\documentclass[conference]{IEEEtran}
\IEEEoverridecommandlockouts
\usepackage{cite}
\usepackage{amsmath,amssymb,amsfonts}
\usepackage{algorithmic}
\usepackage{graphicx}
\usepackage{textcomp}
\usepackage{xcolor}
\usepackage{tikz}
\usepackage{tcolorbox}
\usepackage{enumitem}
\usepackage[breaklinks]{hyperref}

\usetikzlibrary{positioning, arrows.meta, backgrounds}
\UseRawInputEncoding
\def\BibTeX{{\rm B\kern-.05em{\sc i\kern-.025em b}\kern-.08em
    T\kern-.1667em\lower.7ex\hbox{E}\kern-.125emX}}
\begin{document}

\title{Automated Formative Feedback for Short-form Writing: An LLM-Driven Approach and Adoption Analysis}

\author{
    \IEEEauthorblockN{Tiago Fernandes Tavares }
    \IEEEauthorblockA{
        % \textit{} \\
        \textit{Insper}\\
        São Paulo, Brazil\\
        tiagoft1@insper.edu.br
    }
    \and
    \IEEEauthorblockN{Luciano Pereira Soares}
    \IEEEauthorblockA{
        \textit{Insper}\\
        São Paulo, Brazil\\
        lpsoares@insper.edu.br
    }
}    
\maketitle

\begin{abstract}
This paper explores the development and adoption of AI-based formative feedback in the context of biweekly reports in an engineering Capstone program. Each student is required to write a short report detailing their individual accomplishments over the past two weeks, which is then assessed by their advising professor. An LLM-powered tool was developed to provide students with personalized feedback on their draft reports, guiding them toward improved completeness and quality. Usage data across two rounds revealed an initial barrier to adoption, with low engagement rates. However, students who engaged in the AI feedback system demonstrated the ability to use it effectively, leading to improvements in the completeness and quality of their reports. Furthermore, the tool's task-parsing capabilities provided a novel approach to identify potential student organizational tasks and deliverables. The findings suggest initial skepticism toward the tool with a limited adoption within the studied context, however, they also highlight the potential for AI-driven tools to provide students and professors valuable insights and formative support.
\end{abstract}


\section{Introduction}

In the last few years, many fields have been greatly affected by the accelerated evolution and availability of pre-trained large language models (LLMs). LLM-based tools can be used to generate content in various formats, including answering questions, writing programming code, and summarizing research papers. These applications also serve as tutors, like programming assistants or research assistants. However, they can also facilitate advanced forms of cheating, which are becoming increasingly harder to identify~\cite{malinka_educational_2023,herbold_large-scale_2023,rajabi_unleashing_2024}. This duality has inspired a great deal of research that investigates whether LLMs should be adopted or not in the classroom context~\cite{OConnor2023,zou_use_2023,strzelecki_use_2024,hadi_mogavi_chatgpt_2024}

Arguably, education should adapt to newer tools~\cite{hadi_mogavi_chatgpt_2024}, like it has in the past with calculators or with broadly available Internet. However, the misuse of Artificial Intelligence (AI) tools has been shown to hinder learning~\cite{bastani_generative_2024,qureshi_chatgpt_2023}, as they reduce the cognitive effort put into human reasoning and can ultimately make students dependent on the tool's answers. This indicates that there is a need to separate the ability to produce a result with a specific AI tool, and the ability to understand the underlying processes related to that result.

At the same time, the use of AI to provide students with feedback has gained traction in the last few years, particularly after the popularization of LLMs~\cite{shi_systematic_2024,lee_harnessing_2024,huawei_systematic_2023,escalante_ai-generated_2023}. Automated feedback has the advantage of being timely (allowing students to receive instant feedback at any time), but can also be less effective than human feedback~\cite{shi_systematic_2024}. An important feature that has been facilitated by recent LLMs is the ability of providing formative (instead of summative) feedback, that is, feedback that helps students build their knowledge, instead of simply grading their performance.

Even thought AI tools can have a broad range of accuracies, specially when particular problems and domains are considered~\cite{shi_systematic_2024}, their adoption has been shown to be highly related to the users' attitudes~\cite{habibi_chatgpt_2023,almogren_exploring_2024,zou_use_2023,strzelecki_use_2024}. This suggests that the adoption and acceptance of AI tools depend not only on pre-existing propensities unique to each group but also on their role within the learning process~\cite{wambsganss_enhancing_2022}.

% The introduction briefly outlines the background and motivation but does not adequately explain the uniqueness of the Capstone program or the specific writing challenges students face.
In this paper, we discuss the development and adoption of an automated AI-based formative feedback tool in the context of biweekly, low-stakes individual writing tasks within our Capstone program. A Capstone is a culminating academic experience typically undertaken in the final phase of a degree program. In our case˜\cite{soares_cdc_2020, soares_2024_14060955}, it involves students from computer, mechanical, and mechatronics engineering, usually organized into teams of four. As students are required to develop complex projects during the Capstone, professors must continuously monitor their progress, not only through meetings but also through regular written reports. To support this, a two-week cycle was established for students to officially document their progress. Although these biweekly reports are intended to be brief, students often struggle to write clear and objective information that evaluators can use to assess their progress. Their writing frequently includes vague statements that fail to clearly indicate what evidence exists for a given accomplishment.

The AI feedback tool was developed and provided after observing that students often failed to adhere to proper content and form in their biweekly reports, resulting in incomplete reports with structural errors, in other words, low quality reports lacking important information, that could have been avoided through text reviews. These communication issues must be addressed in an educational context, as this underdeveloped skill could lead to career setbacks and project failures in student's  professional future~\cite{Du2019}. The tool was designed to provide formative feedback and reflective questions, but not answers, regarding detected problems in students' texts, as further discussed in Section~\ref{sec:proposed}.

The use of the review AI tool was optional, and students were free to use it as many times as they wanted. This allowed us to evaluate student's perception of the tool usefulness within the context of this specific activity for this research. Since part of the students chose to use the tool while others did not, we were able to compare the results achieved in the reports from both groups. The results, shown in Section~\ref{sec:results}, offer insights into the tool's potential to provide useful feedback, particularly for students who actively engaged with the feedback process.

%We hope the insights revealed in this paper inspire further initiatives on automated feedback and AI-assisted formative assessments.

\section{Proposed tool}
\label{sec:proposed}

%This section is further divided into two subsections. Subsection~\ref{sec:context} explains the context and the course for which the proposed tool was built to. Subsection~\ref{sec:tool} describes the tool itself.

% \subsection{Context}
% \label{sec:context}
The feedback review AI tool was developed for use in the context of an Engineering Capstone program at a private university in Brazil. In this university, Capstone projects are conducted as group work, performed in partnership with companies and industrial partners worldwide. Each group has an advisor professor responsible for assessing students individually and as a group. Students are evaluated based on their final deliverable prototypes, as well as their attitudes and progress throughout the academic semester, which is particularly dependent on their individual participation within the group.

To help advisor professors in this evaluation, every 15 days (approximately), each student must write a brief report in a single field about what they have individually done in the past weeks. The text should be short (a maximum of 2,100 characters). It should not take up too much additional time, neither for the students nor the advisors. This report does not count directly for the final grade, but failure to submit it will result in direct penalties on the final grades.

Students submit their biweekly report in an online, in-house system. They receive instructions as shown in Figure~\ref{fig:student_instructions}.

\begin{figure}[h!]
\begin{tcolorbox}
    Briefly describe YOUR contributions to the project, the knowledge YOU have gained, and the skills YOU have developed during this period. If you wish, you may also include a brief reflection on how you could contribute more effectively to the project in the coming weeks. For each activity performed, provide evidence, such as a created material, a produced drawing, a written code, a researched reference, etc. Additionally, indicate the approximate number of hours you have dedicated to the project since the last biweekly report.
\end{tcolorbox}

\caption{Instructions displayed to students for the biweekly report.}
\label{fig:student_instructions}
\end{figure}

Similarly, advisor professors also receive instructions on using the evaluation system, as shown in Figure~\ref{fig:teacher_instructions}. These instructions comprise not only direct guidance on the procedures for providing feedback but also a discussion of the rationale behind the biweekly reports. This is important because there is a broad diversity of projects within the Capstone program, and a narrowly defined rubric might not be applicable to all cases.

\begin{figure}
\begin{tcolorbox}
The biweekly report aims to provide the advisor with quick feedback on the individual progress of each student in the group, as well as offer the student immediate feedback on their performance as perceived by the advising professor. This continuous feedback seeks to reinforce good practices, identify inappropriate behaviors, highlight areas for improvement, and help the student stay on track for success in the project and approval.

Each student must clearly state their contributions regarding the activities carried out, tasks completed, and decisions made. Students are advised not to use the first-person plural ("we") in these reports, as the focus should be entirely on individual work. Opportunities for feedback and evaluations on group dynamics will be discussed at other situations.

\vspace{0.25cm} \textbf{Optional Feedback}

You may choose to provide additional feedback to students (there is a field for that). If a student is underperforming, it is essential to clearly indicate that their performance was inadequate. While this feedback does not directly impact grades, it is important for the student to recognize poor performance and understand the potential consequences of low grades in the future.

\vspace{0.25cm} \textbf{Evaluating Biweekly Reports}

Consider these points when assessing the report:
\begin{itemize}[left=0.25em]
\item Objectives: Students should describe the goals they set and how much progress they made toward them. Check for clarity and accuracy in the information provided.
\item Planning: Evaluate whether the planning discussed with the advisor is being followed and if any adjustments were made, along with the corresponding justifications.
\item Challenges Faced: Students should report difficulties encountered during the period. Descriptions of technical or logistical problems can provide valuable insights into areas that may need additional support or adjustments.
\end{itemize}
\end{tcolorbox}
\caption{Instructions displayed for advisors while evaluating biweekly reports}
\label{fig:teacher_instructions}
\end{figure}

The problem identified in the biweekly report is that students were writing reports that did not entirely follow the instructions in Figure~\ref{fig:student_instructions}. As a consequence, it frequently became ambiguous whether the difficulties were due to the students' inability to write clearly, or due to the lack of achievements and material to report. In this context, the devised AI tool is oriented for automated formative feedback directed at the biweekly reports, orienting students to improve their text.

\subsection{Automated formative feedback}
\label{sec:tool}
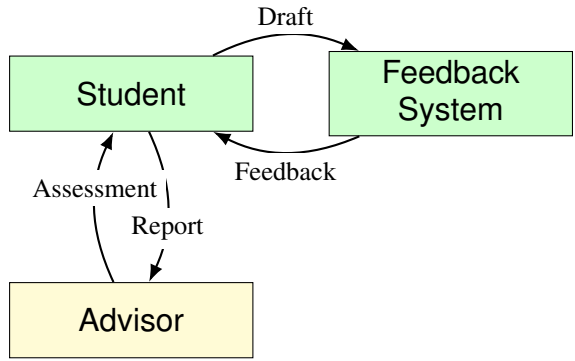
\begin{figure}[h!]
\centering
\begin{tikzpicture}[
  largebox/.style = {
    draw, % Draw the border
    fill=blue!20, % Fill with light blue
    text width=3cm, % Set width
    minimum height=1cm, % Set height
    align=center, % Center the text
    font=\large\sffamily % Larger sans-serif font
  },
  arrow/.style = {-{Latex}, thick, bend left=25}
]

% Nodes
\node[largebox, fill=green!20] (student) {Student};
\node[largebox, fill=green!20, below=2cm of student, right=1cm of student] (system) {Feedback\\System};
\node[largebox, fill=yellow!20, below=2cm of student] (advisor) {Advisor};

% Curved Arrows
\draw[arrow] (student) to[bend left]  node[midway, above, fill=white] {Draft} (system);
\draw[arrow] (system) to[bend left] node[midway, below, fill=white] {Feedback} (student);
\draw[arrow] (student) to node[midway, below, fill=white] {Report} (advisor);
\draw[arrow] (advisor) to node[midway, above, fill=white] {Assessment} (student);
\end{tikzpicture}

\caption{Intended system usage: The student can submit multiple drafts to the AI system and receive feedback on each one. When the student is satisfied with their draft, they can submit it to the advisor for official assessment.}
\label{fig:block_diagram}
\end{figure}

The underlying idea of the proposed AI tool is to enable students to gather formative feedback about their report drafts before submiting a final version to their advisor. This type of usage aims to help students improve their work without needing to wait on an official assessment from their advisor, that may be a request for more information or clear information. Reports that are well-written may enhance the perception of a student's technical work, positively impacting future grades. This process aims to help students write better reports, improving their writing skills and preventing penalties in other competencies due to poor communication. This process is shown in Figure~\ref{fig:block_diagram}.

In this study, the \texttt{gemini-flash}~\cite{gemini} models were utilized. This choice primarily stemmed from a practical reuse of code developed for previous projects. This criterion was used because there is no evidence that current state-of-the-art models are necessarily better than other in a general sense~\cite{lmarena}, thus the choice among them is expected to have a negligible effect in performance.

The system was developed and used in two instances of the biweekly report. Between each of these instances, the system was modified as to incorporate new functionalities. The versions of the system are further described next.

The first developed version for the AI feedback tool used the \texttt{gemini-1.5-flash} model. The prompt used in the system is shown in Figure~\ref{fig:prompt1}. The second version used the (then newly-released) \texttt{gemini-2.0-flash} model. It incorporated some use cases that were not anticipated in the first version. For such, the prompt used is shown in Figure~\ref{fig:prompt2}.

% EXEMPLO
% Over the past two weeks, I focused on integrating the LiDAR sensor with the drone’s onboard processing unit. I successfully established serial communication between the sensor and the Raspberry Pi, allowing real-time data collection. Additionally, I developed a Python script to filter and visualize point cloud data, which will be essential for obstacle detection.  I also coordinated with the laboratory technician to obtain the parts for the sensor mounting.

% One issue encountered was noise in the LiDAR data due to environmental interference. To address this, I researched filtering techniques and implemented a Kalman filter, which significantly improved data stability.

% In the coming weeks, I will work on integrating the obstacle detection algorithm into the drone’s navigation system and begin initial field testing.

Both prompts instruct the system to identify tasks and their corresponding evidences. If no evidence was found, then the task's status should be marked with an error. In the second iteration, more elements could be considered valid evidences, and tasks could be marked as "In Progress".

\begin{figure}[!t]

    \centering
    \includegraphics[width=\linewidth]{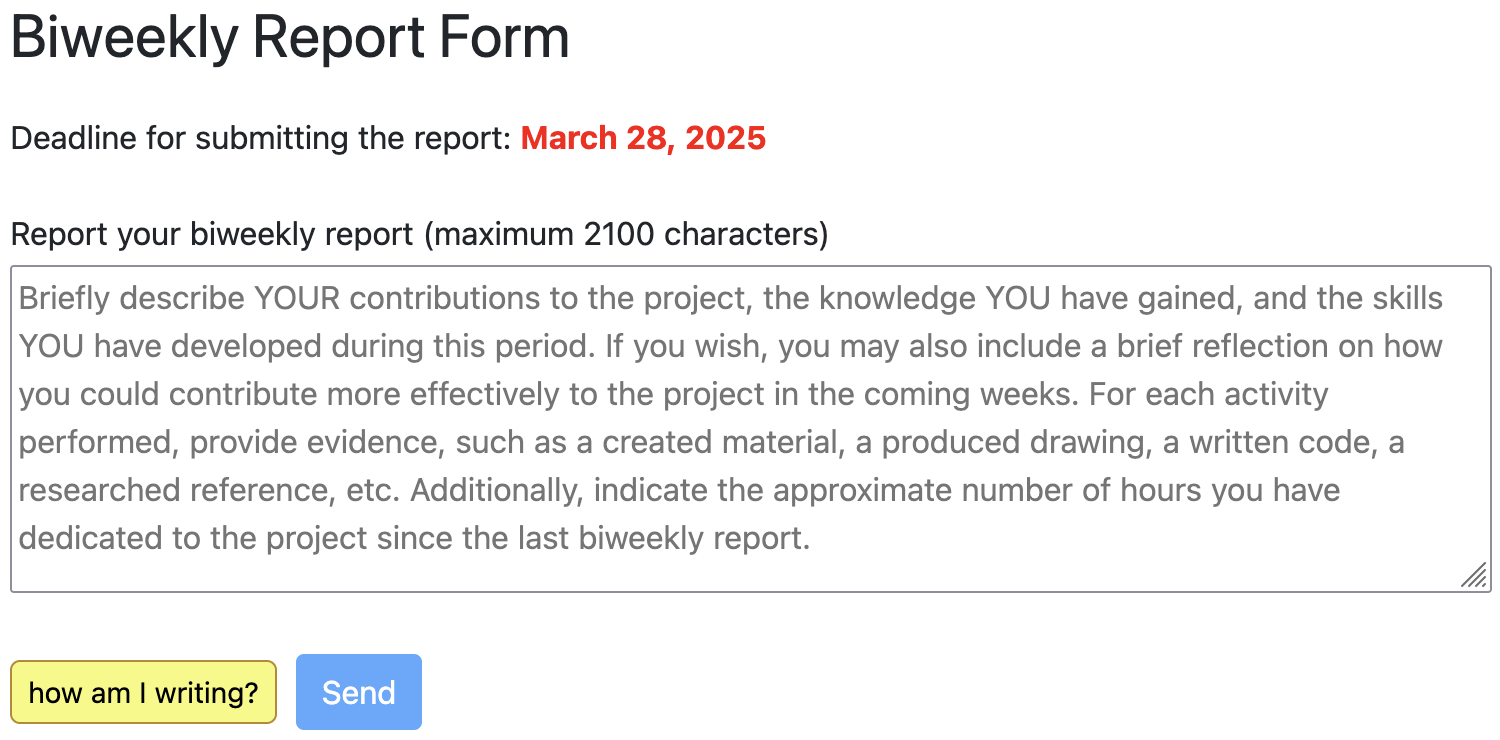}
    \caption{Form used by students to get AI feedback and submit report. The yellow button is used to ask the automatic feedback, and the blue button is used to submit the draft.}
    \label{fig:report_stud_screen}
\end{figure}

\begin{figure}[!t]
    \centering
    \includegraphics[width=\linewidth]{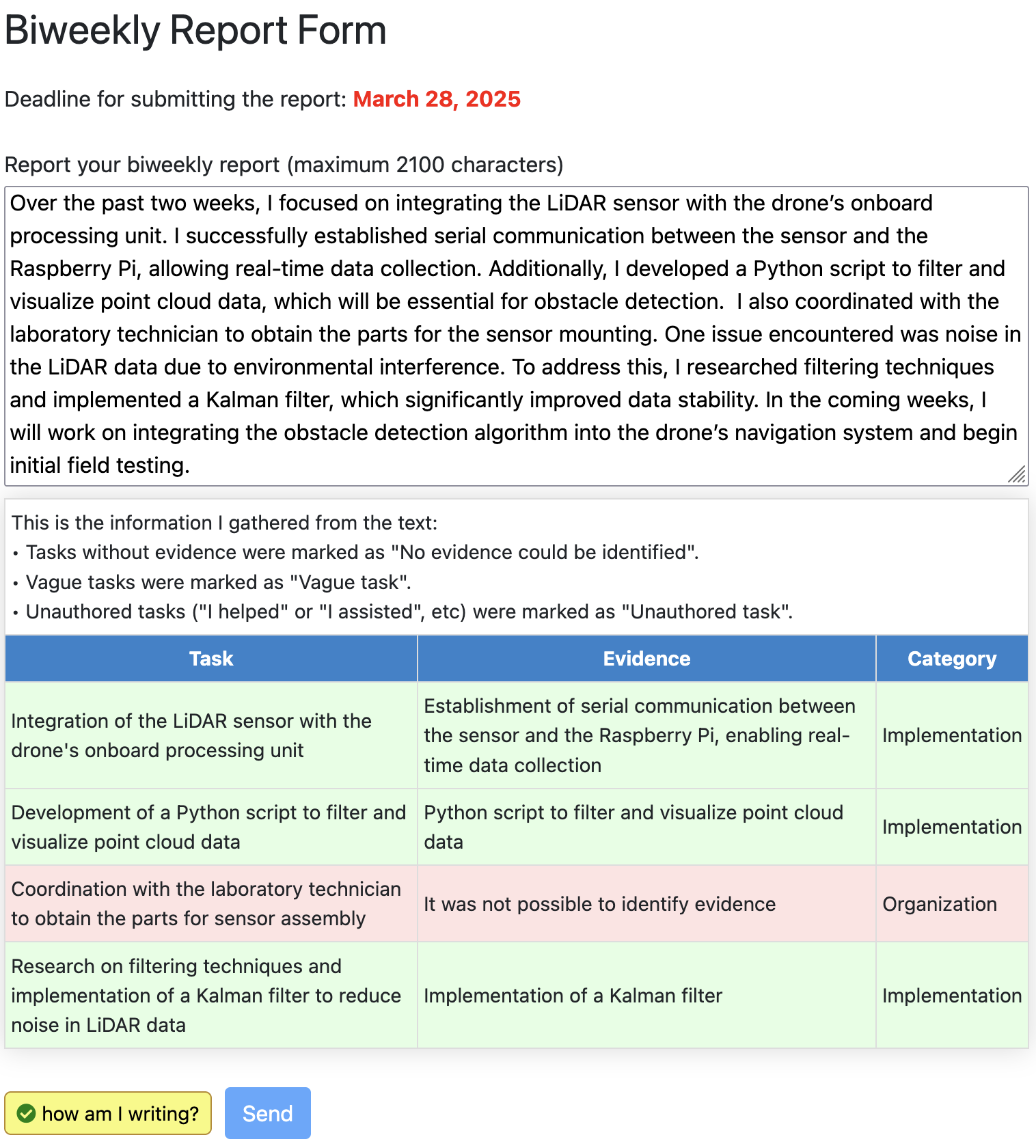}
    \caption{Example of report with the feedback provided. The texts were translated from Portuguese to English.}
    \label{fig:feed_stud_screen}
\end{figure}

The feedback process was incorporated into the submission system so that it could be activated using an auxiliary button, as presented in Figure~\ref{fig:report_stud_screen}. The feedback was shown as a table in which each line corresponded to a task, and the columns correspond to the task description, its evidence, and its status. The lines were colored according to their corresponding status: green for "Ok", red for "Error", and yellow for "In Progress". Figure~\ref{fig:feed_stud_screen} shows an example of report and feedback analysis.

After viewing the feedback, the student can modify their draft and ask for another round of feedback. This can be done as many times as necessary. When the student is satisfied with their draft, they can submit the final version to the advisor.

Importantly, the feedback provided by the system does not present solutions. Instead, it  organizes the main points, and indicates whether each point is clear or requires further refinement. This is meant to leave students with the task of reflecting about and acting upon the information received from the AI.

AI systems can also make mistakes. Because of that, students can ignore the feedback and simply assume that the errors indicated in the feedback are false positives. As planned, the feedback system leaves the decisions and the reasoning to the humans involved in the process.

All student interactions with the AI feedback system were logged, allowing evaluating the adoption and perception of usefulness of the tool by the students. Also, the tool's capability of parsing students activities can be helpful to evaluate students' organization strategies. Data related to both of these aspects is presented on Section~\ref{sec:results}.

\section{Results and discussion}
\label{sec:results}
This section presents the results of our analysis of system usage and student behavior across two rounds of data collection. The platform was offered to Capstone students in the first semester of 2025. All participants were officially in their eighth semester of engineering programs, totaling 76 students: 43 from Computer Engineering, 16 from Mechatronics Engineering, and 17 from Mechanical Engineering. The gender distribution was 13\% women and 87\% men. In the first round 69 of the students submitted the report, on the second round 49 of the students submitted their report.

We examined the frequency of system interactions, student engagement with feedback, and the number of tasks reported by students in their biweekly reports, as detailed in Subsection~\ref{sec:usage_data}. After that, we used the task parsing capability to identify potential organizational problems in students reports, as shown in Subsection~\ref{sec:number_of_tasks}

\subsection{Usage data}
\label{sec:usage_data}
\begin{figure}[ht]
\includegraphics[width=\columnwidth]{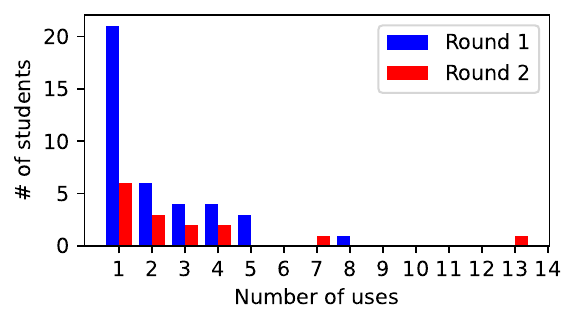}
\caption{Frequency of the number of interactions with the system.
%Most users use the system less than 6 times. Also, there were more users in round 1.
}
\label{fig:usage_frequency}
\end{figure}

Figure \ref{fig:usage_frequency} illustrates the frequency of user interactions with the system. The data reveals that a majority of users engaged with the system relatively few times, with most users interacting only one time. Notably, there were more users in Round 1 compared to Round 2, suggesting a potential decline in user engagement between the two rounds.

\begin{figure}[ht]
\includegraphics[width=\columnwidth]{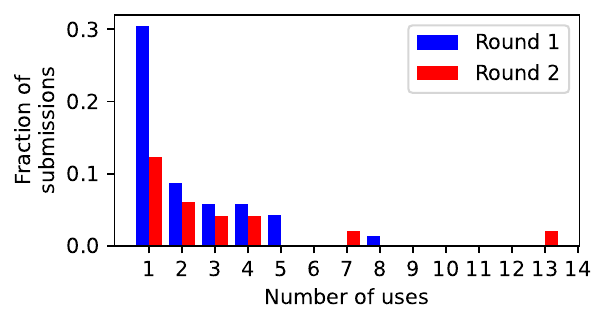}
\caption{Frequency of the number of interactions with the system, relative to the number of submissions. Even in this case, round 2 had less users than round 1.}
\label{fig:usage_frequency_rel}
\end{figure}

Accounting for varying numbers of submissions per round, Figure \ref{fig:usage_frequency_rel} presents the frequency of system interactions normalized by the number of submissions. Despite this adjustment, the trend remains consistent with Figure \ref{fig:usage_frequency}: Round 2 consistently exhibited fewer users than Round 1. This suggests that the reduced user base in Round 2 was not just due to fewer submissions; rather, it consists of a genuine decline in the proportion of students actively using the system.

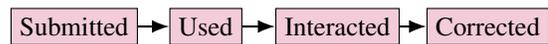
\begin{figure}[h!]
\centering
\begin{tikzpicture}[
  largebox/.style = {
    draw, % Draw the border
    fill=blue!20, % Fill with light blue
    text width=3cm, % Set width
    minimum height=1cm, % Set height
    align=center, % Center the text
    font=\large\sffamily % Larger sans-serif font
  },
  arrow/.style = {-{Latex}, thick, bend left=25}
]

% Nodes
\node[draw, fill=purple!20] (submitted) {Submitted};
\node[draw, fill=purple!20, right of=submitted, xshift=0.75cm] (used) {Used};
\node[draw, fill=purple!20, right of=used, xshift=0.75cm] (interacted) {Interacted};
\node[draw, fill=purple!20, right of=interacted, xshift=1cm] (corrected) {Corrected};

% Curved Arrows
\draw[arrow] (submitted) -- (used);
\draw[arrow] (used) -- (interacted);
\draw[arrow] (interacted) -- (corrected);
\end{tikzpicture}
\caption{User trajectory illustrating key  actions: submit a report, use the system at least once, interact with the system at least twice, and use the system's feedback to improve drafts. As expect, the number of users decreased at each step, thus forming a usage funnel.}
\label{fig:diagram_usage_funnel}
\end{figure}

We further analyze user trajectories while interacting with the system. Such analysis assumes, as depicted in Figure~\ref{fig:diagram_usage_funnel}, that users that submit the report can use the system, and then a fraction of those that use the system can interact twice or more with it, and, finally, that a fraction of the users who interacted with the system actually used its feedback to improve their drafts. Users were considered to have improved their drafts when their last use of the feedback system pointed less errors (as detected by the feedback system) than their corresponding first attempt.

\begin{figure}[ht]
\includegraphics[width=\columnwidth]{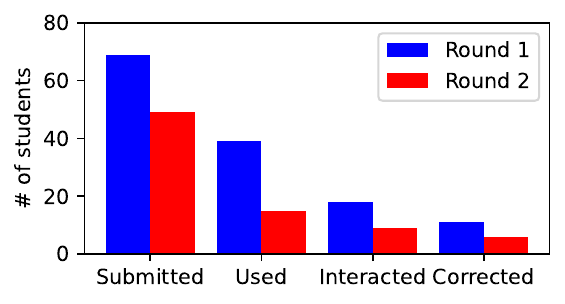}
\caption{Usage funnel displaying the number of students who submitted the report, used the system, interacted with it more than once, and used feedback to improve their texts. The usage reveals a progressive decline in engagement.}
\label{fig:usage_funnel}
\end{figure}

Usage data showing the progressive disengagement with the system at various points is depicted in Figure \ref{fig:usage_funnel}. Clearly, there is a progressive decline in engagement through the user steps, which culminates with only a few students (from the total cohort) having actually corrected their drafts using the feedback system. However, we observe that the number of students that the relative declines (i.e., the fraction of students that continued using the system) from one step, that is, the relative attrition, to another changes significantly.

\begin{figure}[ht]
\includegraphics[width=\columnwidth]{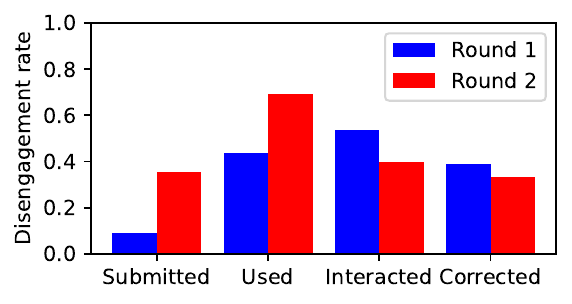}
\caption{Disengagement rates for each step in the usage funnel. Numbers shown are relative to the previous step, e.g., the bar in ``interacted'' shows the fraction of users that disengaged with the system after the ``used'' step. The highest disengagement rate was in the ``interacted'' step in Round 1, but was in the ``used'' step in Round 2.}
\label{fig:usage_funnel_rel}
\end{figure}

The relative attrition between each stage of the funnel is brought forward in Figure \ref{fig:usage_funnel_rel}. The data reveals that the highest disengagement rates occur in the first interaction steps (``used'' and ``interacted''). However, users that engage in interacting with the system frequently use it for corrections.

Usage data suggests an initial barrier to start engaging with the draft-feedback loops (as shown in Figure~\ref{fig:block_diagram}). However, students who engage with the feedback process tend to use it correctly and benefit from it. In both rounds, over half of the students who interacted with the system could improve their drafts (according to the system's feedback).

However, we note that this particular group of students exhibited significant disengagement from the activity itself, that is, many students did not submit the report in Round 2. Hence, disengagement with the feedback tool could be a part of this more general avoidance towards the task of submitting regular reports. This could be seen as an opportunity to review and discuss students' attitudes towards the task of regularly submitting reports.

\subsection{Observing student organization}
\label{sec:number_of_tasks}

The feedback tool can help advisors and the capstone coordinators to identify possible problems in student organization, for example by observing the number of tasks reported by each student. Too few or too many tasks reported by the AI tool can indicate that students are misunderstanding the reporting proposal or are inaccurate (under- or over-reporting) in their task tracking.

\begin{figure}[ht]
\includegraphics[width=\columnwidth]{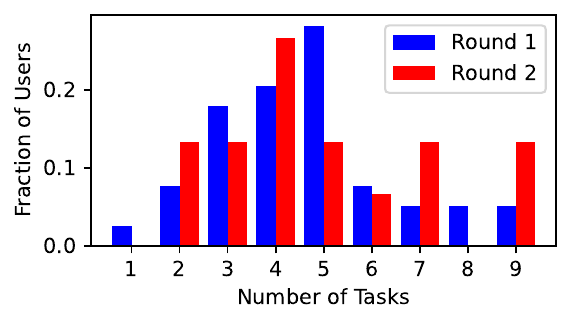}
\caption{Number of Capstone project tasks reported by students in the biweekly report. In general, students perform around 4 to 5 tasks. However, we can observe that some students report too many (more than 8) or too few (1) tasks, which can indicate some problem.}
\label{fig:number_of_tasks}
\end{figure}

Figure \ref{fig:number_of_tasks} presents the distribution of the number of tasks reported by students in their biweekly reports. The data indicates that students generally reported performing between four to five tasks. However, these data also highlight the presence of outliers, with some students reporting either excessively high (more than eight) or low (one) numbers of tasks. Advisors can use this data in combination with other information to officially give a feedback to the student.

\begin{figure}[ht]
\includegraphics[width=\columnwidth]{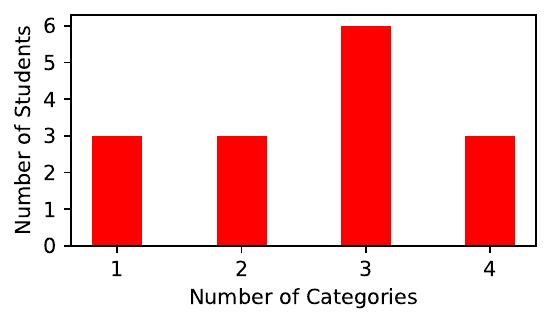}
\caption{Number of task categories found in student biweekly reports in Round 2. Although the number of reported tasks spans up to 9, these tasks are, in fact, condensed into up to 4 categories.}
\label{fig:number_of_categories}
\end{figure}

Also, the automatic system can help highlighting student preferences for particular types of tasks. As shown in Figure~\ref{fig:prompt2}, the second version of the system (applied in round 2) is able to categorize tasks. This allows two analyze the number of task categories performed by each student, shown in Figure~\ref{fig:number_of_categories},

The classification of tasks into categories can help identifying students that focus too much in a single task category (for example, students that only ``write reports'' and ``participate on meetings''). Further, these observation could help diagnose group organization issues, such as students that consistently perform fewer or narrower tasks than their colleagues. Such an idea requires a longer observation time, thus it is out of scope for this paper.

%The next section brings conclusive remarks.

\section{Conclusion}
In this paper, we present the development of an LLM-based formative feedback tool and investigated the adoption of an within the context of a Capstone project's biweekly reports. Usage data reveals a notable initial barrier to entry despite the potential for improving students' report drafts. The tool's ability to parse student tasks also offered a novel perspective to identify of potential organizational challenges within working groups.

The developed AI tool is focused on formative feedback and does not present ready-made answers. As such, it does not assume the role of the instructor in grading, which could raise ethical concerns, and it does not drastically lower the cognitive load for students, which could negatively impact learning. Rather, it offers a timely guidance by providing suggestions that students can choose to follow or ignored.

Additionally, we identify that LLMs can help analyzing student reports using statistical approaches. This could highlight pathways for more effective and measurable interventions.

The results shown in this work are limited to particular activity within a specific Capstone program cohort. However, the system's core components -- LLM-based text analysis, task-evidence mapping, and personalized feedback generation -- can be adapted for different subject areas, writing genres, and assessment types. Thus, while this study focused on biweekly reports, the system could be extended to provide feedback on other forms of student writing, such as essays, research proposals, or technical documentation. These adaptations may yield different implications and takeaways, particularly regarding the adoption funnel. Such investigations will be pursued in future work.

\bibliographystyle{IEEEtran}

\bibliography{references}
\begin{figure*}[!ht]
\begin{tcolorbox}
You have the task of analyzing a report from an engineering student. This report comprises a biweekly summary of the activities carried out by the student. Your task is to identify the tasks performed by the student and the material evidence the student claims to have produced. We would like the report to always include, as content, the actions and tasks carried out by the student, as well as evidence showing that the task was performed. The evidence must be a reference to some material produced by the student, such as a code, a report, a link, etc., that can be accessed. A study or research is not valid evidence. References to texts and tables are valid evidence.

Return your response in the following JSON format:

\begin{verbatim}
{
  "tasks": [
    {
      "Task": "Description of task 1",
      "Evidence": "Evidence of task 1",
      "Status": "OK"
    },
    {
      "Task": "Description of task 2",
      "Evidence": "Evidence of task 2",
      "Status": "OK"
    },
    {
      "Task": "Description of task N",
      "Evidence": "Evidence of task N",
      "Status": "OK"
    }
  ]
}
\end{verbatim}
The task is to find information, not to interpret the text.

If you cannot identify any tasks, return an empty list.

If you cannot identify evidence (a link or a reference to some material) for any task, or if the evidence is not a link or a reference to material produced by the student, mark the field "Evidence" as "No evidence could be identified" and mark the status as "ERROR."

If any task is vague, mark the field "Task" as "Vague task" and mark the status as "ERROR."

If any task is identified without the authorship of the student, such as "I helped," "I participated," "I assisted," mark the field "Task" with the description of the task and add the tag "(Unauthored task)" and mark the status as "ERROR."
\end{tcolorbox}

\caption{System prompt used in the first iteration of the system. The prompt requires an explicit link to student material to consider a task as "correct".}
\label{fig:prompt1}
\end{figure*}

\begin{figure*}[ht]
\begin{tcolorbox}[colback=blue!10]
You have the task of analyzing a report from an engineering student. This report consists of a biweekly summary of activities carried out by the student. Your task is to identify the tasks performed by the student and the material evidence the student claims to have produced.

Return your response in a JSON format as follows:

\begin{verbatim}

{
  "tasks": [
    {
      "Task": "Description of task 1",
      "Evidence": "Evidence of task 1",
      "Category": "Category of task 1",
      "Status": "OK"
    },
    {
      "Task": "Description of task 2",
      "Evidence": "Evidence of task 2",
      "Category": "Category of task 2",
      "Status": "OK"
    },
    {
      "Task": "Description of task N",
      "Evidence": "Evidence of task N",
      "Category": "Category of task N",
      "Status": "OK"
    }
  ]
}
\end{verbatim}

The task is solely to find information and not to interpret the text.

If the task involves studying, research, or initial testing, mark the category as "Study."

If the task involves implementation, development, prototyping, machining, assembling, etc., mark the category as "Implementation."

If the task involves writing reports, mark the category as "Writing."

If the task involves organizing the group, contacting people, scheduling meetings, etc., mark the category as "Organization."

If the task involves attending meetings, mark the category as "Meeting."

For "study"-type tasks, valid evidence includes links to materials produced by the student, direct mentions of what was learned in the study, or the results of the tests. For "implementation"-type tasks, valid evidence includes links or mentions of codes, prototypes, assemblies, etc. For "writing"-type tasks, valid evidence includes links or mentions of reports, texts, repositories, etc. For "organization"-type tasks, valid evidence includes links or mentions of meeting minutes, decisions made, messages sent, etc. For "meeting"-type tasks, valid evidence sought includes decisions made during the meetings, direct mentions of what was discussed, or direct mentions of what was presented by the student.

If a task has no evidence, mark the status as "ERROR" and the evidence as "No evidence could be identified."

If a task is still in progress, mark the status as "IN PROGRESS" and the evidence as "Task in progress."

If no tasks can be identified, return an empty list.

If any task is vague, meaning it is not possible to identify what was done by the student, mark the field "Task" as "(Vague task: be specific about what was done)" and mark the status as "ERROR."

If any task is identified without being authored by the student, or is attributed to the group as a whole ("we did," "the group did," "member X, Y, Z did"), or is attributed as "I helped" or "I assisted," mark the "Task" field with the description of the task but add the tag "(Unauthored task: mention only your own actions)" and mark the status as "ERROR."
\end{tcolorbox}

\caption{System prompt used in the second iteration of the system. The system is now able to identify more self-evident tasks, such as studies that are evident due to mentions to what was learned (instead of a link to an online report), can divide tasks into categories, and is able to identify ongoing tasks.}
\label{fig:prompt2}
\end{figure*}
\end{document}